\documentclass[aps,nofootinbib,showpacs,twocolumn]{revtex4-1}
\usepackage[CJKbookmarks, pdftex, bookmarksnumbered, bookmarksopen, colorlinks, citecolor=blue, linkcolor=blue]{hyperref}
\usepackage[english]{babel}
\usepackage{amstext,amsmath,amssymb,amsfonts,bbm}
\usepackage[latin1]{inputenc}
\usepackage{graphicx}
\usepackage{epstopdf}
\usepackage{amsthm}
\usepackage{tocvsec2}
\usepackage{enumerate}
\usepackage{subfigure}
\usepackage{color}
\usepackage{tikz}
\usepackage[squaren,Gray]{SIunits}

\newcommand{\va}{\scriptscriptstyle}

\newcommand{\be}{\nopagebreak[3]\begin{equation}}
\newcommand{\ee}{\end{equation}}

\newcommand{\bee}{\nopagebreak[3]\begin{equation*}}
\newcommand{\eee}{\end{equation*}}

\newcommand{\ba}{\nopagebreak[3]\begin{eqnarray}}
\newcommand{\ea}{\end{eqnarray}}

\begin{document}

\title{Dark energy as the weight of violating energy conservation}

\author{Thibaut Josset, Alejandro Perez}
\affiliation{{Aix Marseille Univ, Universit\'e de Toulon, CNRS, CPT, Marseille, France}}

\author{Daniel Sudarsky}
\affiliation{Instituto de Ciencias Nucleares, Universidad Nacional Aut\'onoma de M\'exico, M\'exico D.F. 04510, M\'exico}

\date{November 21, 2016}

\begin{abstract}
	In this letter, we consider the possibility  of reconciling metric theories of gravitation with violation of the conservation of energy-momentum. Under some circumstances, this can be achieved in the context of unimodular gravity, and it leads to the emergence of an effective cosmological constant in Einstein's equation.   We  specifically investigate   two potential sources of energy non-conservation---non-unitary modifications of quantum mechanics,  and phenomenological models motivated by quantum gravity theories with spacetime discreteness at the Planck scale---and show that such locally negligible phenomena can nevertheless become relevant at the cosmological scale.
\end{abstract}

\maketitle

	Ever since the  discovery of the   acceleration  in the  universe's   expansion  \cite{Riess:1998cb, Perlmutter:1998np},   almost  two decades  ago,  there   has  been a puzzlement about the  strange  value of the corresponding  cosmological constant $\Lambda$;  the simplest, and so far most successful, theoretical model that could account for the observed behaviour.   The origin of this puzzle is that, within the usual framework, the only seemingly natural values that $\Lambda$ could take are  either zero, or a  value  which  is 120  orders of magnitude  larger than the one indicated by observations $\Lambda^\text{obs} \approx 1.1~10^{-52}~\meter^{-2}$  \cite{Weinberg:1988cp}.

	In this letter we present a scenario where something very similar to  a cosmological constant  emerges from  certain sources  of  violations of  energy-momentum  conservation and their influence   on the space-time  geometry.  We  will   consider  here  two different sets of ideas,   motivating such   violations,  but  these could also be introduced at a purely phenomenological level. On the one hand, such violations are commonplace in the context of certain non-unitary modifications  of Schr\"odinger's equation  \cite{Bassi:2012bg} proposed as a way to  address  the measurement problem in quantum theory \cite{Meassurment}. 
	On the other hand, they are natural in quantum gravity approaches where fundamental spacetime discreteness could lead to small violations of translational invariance. A concrete model of phase space diffusion due to Planckian granularity (proposed in the context of causal sets in \cite{Dowker:2003hb,Philpott:2008vd}) will be used here as an example.

	One of   the most  serious difficulties faced  by  such proposals  relates to their  consistency  (or lack  thereof) with the gravitational interaction. We point out that this tension  can  be  resolved in the framework of unimodular gravity in the cosmological setting: this resolution  leads to  the appearance of an effective cosmological constant  that registers the  cumulative effect  of the  lack of energy-momentum conservation. In the cosmological setting, we estimate the contribution to the effective cosmological constant arising from violations of energy-momentum conservation predicted by modified quantum mechanics models as well as by the proposal based  on  the causal set approach  to quantum gravity. We show that these contributions  can be comparable in size with the value of the cosmological constant inferred from current observations.

	Thus, and on  a more general ground,  our work proposes a new paradigm for analyzing the dark energy puzzle in cosmology, that identifies potential violations of energy-momentum conservation in the past (that could  be postulated on a simply phenomenological ground) as a source of dark energy today.
\\  
 
	In general relativity, local energy-momentum conservation, $\nabla^b  \langle T_{ab}\rangle =0$, is a consequence of the field equations, both at classical and semi-classical levels. This is obvious from the semi-classical version of Einstein's equation 
\be \label{EinsteinEq}
	R_{ab} - \frac{1}{2} R g_{ab}   =  \frac{8 \pi G}{c^4}   \langle T_{ab}\rangle
\ee
---where  $\langle T_{ab}\rangle $ is the expectation value of the   (renormalized)  energy-momentum  tensor operator  in the corresponding  quantum state of the matter fields---and the fact that the Bianchi identities make the  geometric side divergence free\footnote{For  other  views on the  issue we refer the reader to the arguments  claiming that semi-classical  GR is  simply  unviable  \cite{Page-Geilker}, a dissenting opinion \cite{Carlip},  and for  an alternative way of looking  at such theory \cite{nosotros SSC}.}.

	The previous restriction can be circumvented by considering a simple modification of general relativity, already evoked by Einstein in 1919 when trying to construct a geometric account for elementary particles   in terms of radiation fields \cite{Einstein1919Spielen-Gravita}.  He  proposed the trace-free equation 
\begin{equation}\label{TraceFreeEinsteinEquation}
	R_{ab} - \frac{1}{4} R g_{ab} = \frac{8 \pi G}{c^4} \left( T_{ab} - \frac{1}{4} T g_{ab} \right),
\end{equation}
which has been rediscovered several times, and is now called unimodular gravity (see \cite{Ellis:2010uc} and references therein). Unimodular gravity can be derived from the Einstein-Hilbert action by restricting to variations preserving the volume-form, i.e., those for which $g_{ab}\delta g^{ab}\!=0$. 

This breaks the diffeomorphism symmetry down to volume-preserving 
diffeomorphism, whose infinitesimal version is given by divergence-free vector fields $\xi^a$, i.e., 
\be\label{ff}
\nabla_a\xi^a=0.
\ee
This restriction on general covariance allows for violations of energy-momentum conservation of a  certain form. To see this, consider an action for matter $S_m$ invariant under volume-preserving diffeomorphisms, introduce the stress-energy tensor $T_{ab} \equiv - 2|g|^{-1/2}{\delta S_m}/{\delta g^{ab}}$, and its energy-momentum violation current $J_a\equiv \nabla^bT_{ab}$. The variation of the action under an infinitesimal diffeomorphism (of compact support) $\xi^a$  is 
\ba\label{variation-of-the-action}
\delta S_m = - \int T_{ab} \nabla^a \xi^b \sqrt{-g}dx^4 = \int  J_a\xi^a \sqrt{-g}dx^4 ,
\ea 
where the matter fields equations are assumed to hold.  Inserting the general solution of  \eqref{ff}  $\xi^a=\epsilon^{abcd}\nabla_b \omega_{cd}$---for an arbitrary two-form $\omega$---the requirement that the action is invariant under volume-preserving diffeomorphisms  ($\delta S_m=0$) implies $dJ=0$. Hence violations of energy-momentum conservation are allowed  in unimodular gravity  as long as they are of such integrable type.

For simply-connected spacetimes, this condition reduces to  \be \label{IntegrableType} J_a =\nabla_aQ,\ee for some scalar field $Q$. Thus, if the matter action is only invariant under volume-preserving diffeomorphisms, then $J\not=0$ will introduce deviations from general relativity. We will discuss later the naturalness of such symmetry breaking in quantum field theory.

	An important feature of unimodular gravity  in the semi-classical framework is that vacuum fluctuations of the energy-momentum tensor do not gravitate \cite{Weinberg:1988cp}.   This removes the need to contemplate the enormous discrepancy   between the  observed  value of the cosmological constant,  and  the   standard  estimates  from  the  vacuum  energy  \cite{Ng:1990xz, Smolin:2009ti,Ellis:2010uc}.\footnote{It has also been argued that unimodular gravity may not suffer from the problem of time \cite{Unruh:1988in, Henneaux:1989zc}; however, this view has been criticized and clarified in \cite{Kuchar1991}.}

	Let us move on and  consider  the semi-classical  version  of  equation (\ref{TraceFreeEinsteinEquation}),  where  the  energy-momentum tensor and its   trace are now replaced by the  corresponding expectation values  in a  quantum state  of the matter fields.  Using Bianchi identities, one then  deduces that
\begin{equation}\label{BianchiUG}
	\frac{1}{4} \nabla_a R = \frac{8 \pi G}{c^4} \left( \nabla ^b \langle T_{ab}\rangle - \frac{1}{4} \nabla_a \langle T \rangle \right),
\end{equation}
which, after integration, can be used to recast  \eqref{TraceFreeEinsteinEquation}  as
\begin{equation}\label{TraceFreeEinsteinEquation2}
	R_{ab} - \frac{1}{2} R g_{ab} +\left(\Lambda_{\va -\infty} +\frac{8 \pi G}{c^4} Q
	\right) g_{ab}= \frac{8 \pi G}{c^4} \langle T_{ab} \rangle ,
\end{equation}
where $\Lambda_{\va -\infty}$ is a constant of integration, and $Q$ is defined by \eqref{IntegrableType}  
\footnote{Of course, the equation of motion \eqref{TraceFreeEinsteinEquation2} derived from unimodular gravity is completely equivalent to the use of the conserved stress-energy tensor $\tilde T_{ab} \equiv T_{ab}-Q g_{ab}$ in the Einstein equations. In both cases, to make sense of $Q$ as a local quantity, the integrability condition needs to be satisfied.}. As expected, when the stress-energy tensor is conserved, i.e., $Q=0$, \eqref{TraceFreeEinsteinEquation2} simply reduces to  Einstein's equation, with a cosmological constant equal to $\Lambda_{\va -\infty}$.

We emphasize that both semiclassical general relativity and its unimodular version are regarded here as an effective and emergent description of more fundamental degrees of freedom (just like the Navier-Stokes description of a fluid). The violation of energy-momentum conservation, in our scenario would have to admit a description in terms of the more fundamental, presumably,  quantum gravity degrees of freedom.

	Specializing to cosmology, and considering an homogeneous, isotropic, and spatially flat Friedmann-Lema\^itre-Robertson-Walker universe, $ds^2 = -c^2dt^2+a^2 d\vec x^2$,  the modified Friedmann equation reads
\be
	H^2 \equiv \left( \frac{\dot a}{a}\right)^2 = \frac{8\pi G}{3 c^2}\rho(t) + \frac{\Lambda^\text{eff}(t) c^2}{3} ,
\ee
where the effective cosmological ``constant''
\begin{equation}\label{Eff-Lambda}
	\Lambda^\text{eff}(t)  \equiv \Lambda_{\va -\infty} + \frac{8 \pi G}{c^4} \int^t J
\end{equation}
registers the possible violations of energy-momentum conservation in the past history of the universe. We have re-expressed $Q=\int^t J$ as it will be more convenient for explicit calculations in the following paragraphs. As we shall see later, small violations of energy-momentum conservation---that might remain inaccessible to current tests of local physics---can nevertheless have important cosmological effects at late times, in the form of a nontrivial contribution to the present value of the cosmological constant. 
\\


	The first scenario---leading to violation of energy-momentum conservation---that we explore is the one offered by non-unitary modifications of quantum dynamics. In  order to recover Born's rule for probabilities of experimental outcomes, these modifications of quantum theory involve non-linearity and stochasticity, which for a wide class of models can be described by a Markovian evolution equation for the density matrix $\hat \rho$: the so-called Kossakowski-Lindblad equation  \cite{Kossakowski1972On-quantum-stat,Lindblad:1975ef}
 \begin{equation}\label{ModifEvol}
 	\dot{\hat \rho}  = - i [\hat H,  \hat \rho] - \frac{1}{2} \sum_{\alpha} \lambda_{\alpha}  [\hat K_{\alpha}, [\hat K_{\alpha}, \hat  \rho]]~,
\end{equation}
  where   $\hat H$  is the standard Schr\"odinger   Hamiltonian  operator,  $\lbrace \hat K_{\alpha}\rbrace $  are hermitian operators  characterizing the modified  dynamics,  and  $\lbrace \lambda_{\alpha}\rbrace $  are suitable parameters determining the  strength of the new effects. 

	Such equation has been used to describe a possible non-unitary evolution induced by the creation and evaporation of black holes   \cite{Peskin, Wald-Unruh},  in the context of Hawking's information puzzle  \cite{Hawking:1976ra}. It also appears in the description of  modifications of quantum mechanics with spontaneous stochastic collapse \cite{Bassi:2012bg}. 
It has been argued by Penrose  \cite{Penrose:1996cv} that the two apparently different contexts could actually be related in a more fundamental description of quantum gravitational phenomena (for a recent development see \cite{A-Tilloy}). Finally, the previous equation would also arise in the description of decoherence with underlying discrete spacetime \cite{Perez:2014xca, Gambini:2004qx, Gambini:2004de}. In all these cases, a generic feature of equation \eqref{ModifEvol} is that the average energy $\langle E \rangle \equiv {\rm Tr}[\hat \rho \hat H]$ is not constant.

\begin{figure}
	\includegraphics[scale=0.65]{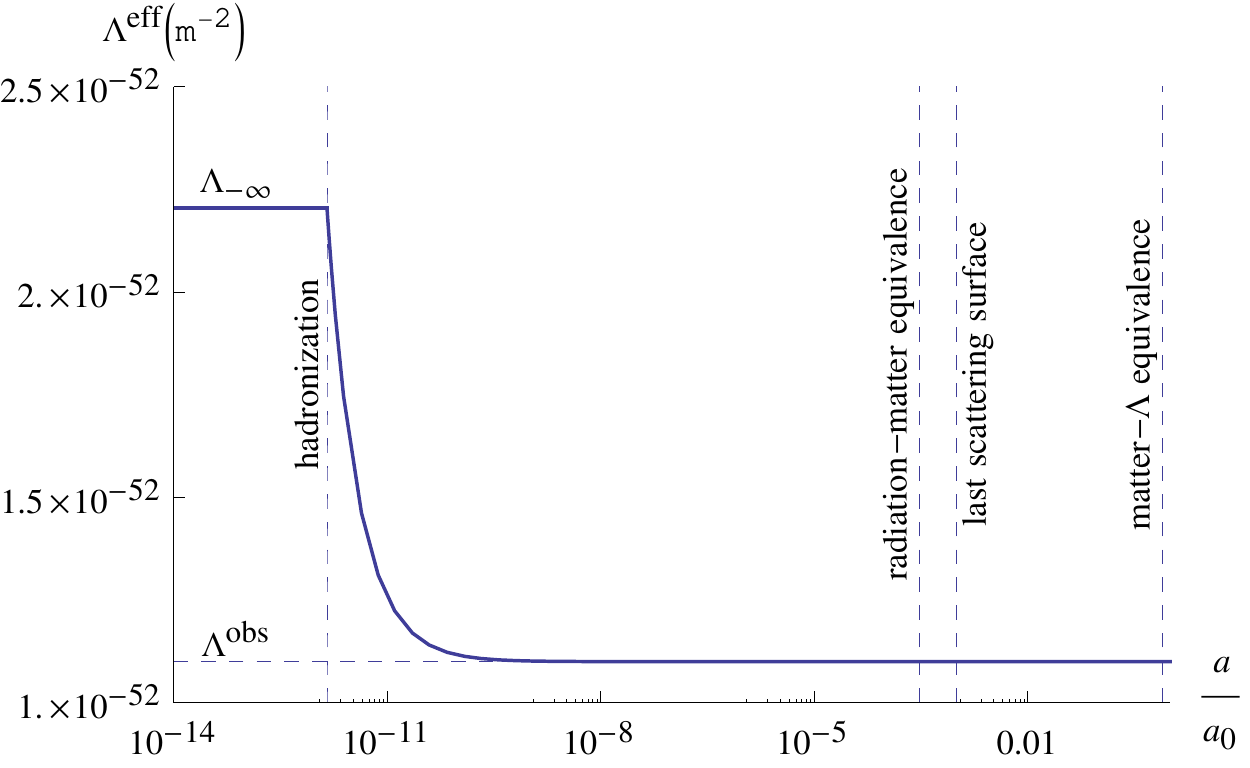}
	\caption{Effective cosmological constant induced by wave-function collapse of baryons, using mass-proportional CSL model with 
	$\xi_{\va \text{CSL}} =4.3 \,10^{-31}\, \second^{-1}$.}
	\label{solveDE}
\end{figure}

	One of the prominent models of this type, for non-relativistic particles, is the so-called mass-proportional continuous spontaneous localization (CSL) model \cite{Pearle1976Reduction-of-th, Ghirardi:1985mt, Pearle1989Combining-stoch, Ghirardi1990Markov-processe}, obtained when $\hat K_\alpha$ are smeared mass-density operators. It exhibits a ceaseless creation of energy proportional to the mass of the object collapsing \cite{Pearle1994Bound-state-exc}. Thus, in the cosmological context,  the CSL of baryons leads to an energy-momentum violation current 
\be\label{EnergyIncreaseCSL}
	J = -\xi_{\va \text{CSL}} \rho^\text{b} dt,
\ee
where $\rho^\text{b}$ is the energy-density of the baryonic fluid, and the parameter $\xi_{\va \text{CSL}}$ is constrained by current experiments according to $3.3\,10^{-42} \second^{-1} < \xi_{\va \text{CSL}} < 2.8\,10^{-29} \second^{-1}$ (see figure 4 in \cite{Bassi-figure}). Choosing hadronization ($z_{\rm h}\approx 7 \,10^{11}$) as initial time, \eqref{Eff-Lambda} and \eqref{EnergyIncreaseCSL} lead to
\begin{equation}\label{DarkEnergyEstimate1}
	\Delta \Lambda^\text{eff}_{\va \text{CSL}} \approx -  \frac{3 \Omega^\text{b}_0 H_0 \xi_{\va \text{CSL}}}{\sqrt {\Omega^\text{r}_0}c^2}  z_{\rm h} \approx   -  \frac{\xi_{\va \text{CSL}}}{4.3\,10^{-31}\, \second^{-1}}  \Lambda^\text{obs},
\end{equation}
where $ \Lambda^\text{obs}$ is the observed value of the cosmological constant, and we used standard values for the cosmological parameters \cite{Adam:2015rua}\footnote{For simplicity, the contribution to $\Omega^r$ of particles like electrons, muons or pions, which were relativistic at the hadronization epoch has been neglected. This would affect the estimate \eqref{DarkEnergyEstimate1} by a numerical factor of order 1.}.
As the effect is linear in the matter density, $\Lambda_{\rm eff}$ becomes quickly a constant (figure \ref{solveDE}).
\\

	The second scenario where  violations of energy-momentum conservation have been argued to  arise naturally is the causal set approach to quantum gravity \cite{Dowker:2003hb, Philpott:2008vd}. These effects are shown to be  compatible with Lorentz invariance, they are  described for both massive and massless particles, and controlled by a few phenomenological parameters.  More precisely, for free massless particles, the physics is encoded in a phase space diffusion equation that  reads
\be
	\frac{d\mu}{dt}=-\frac{p^i}{E}\partial_i \mu - \left(k_1+k_2 \right) \frac{\partial \mu}{\partial E} + k_1 E \frac{\partial^2 \mu}{\partial E^2},
\ee
where $k_1$ and $k_2$ have been constrained comparing the CMB with Planck's spectrum \cite{Philpott:2008vd}. In the cosmological context, the diffusion in phase space leads to an energy-momentum violation current of the form
\be \label{EnergyCreationCS}
	J= -(3k_1+k_2) n^\gamma dt =- \xi_{\va {\va \text{CS}}} \rho^\gamma_0 \left(\frac{a_0}{a}\right)^3 dt
\ee
 where $n^\gamma$ is the number-density of photons, and $-10^{-21} \, \second^{-1}< \xi_{\va {\va \text{CS}}}< 2\, 10^{-21} \, \second^{-1} $. Interestingly, $\xi_{\va {\va \text{CS}}} $ can be negative (endothermic evolution), and thus contributes positively to the effective cosmological constant. Being very conservative, we can estimate that contribution starting from when photons decoupled from electrons ($z_{\rm dec}\approx 1100$); the result is
\begin{equation}\label{DarkEnergyEstimate2}
	\Delta \Lambda^\text{eff}_{\va \text{CS}} \approx - \frac{2  \Omega^{\gamma}_0H_0 \xi_{\va \text{CS}}}{\sqrt {\Omega^\text{m}_0}c^2} z_{\rm dec}^{{3}/{2}} \approx- \frac{\xi_{\va \text{CS}}}{6 \, 10^{-19}\, \second^{-1}} \Lambda^\text{obs}.
\end{equation}
\\

	Both results  \eqref{DarkEnergyEstimate1} and \eqref{DarkEnergyEstimate2} are very sensitive to the initial time at which violations of energy conservation started. In the case of CSL, a precise description of the quark-gluon plasma to hadron gas transition is difficult \cite{Hadronization}; we simply assumed it to be instantaneous at $T_\text{QCD} = 2\, 10^{12} \, \kelvin$. Moreover, from an objective collapse perspective, one also expects modifications of quantum mechanics for relativistic particles and interacting systems but, due to the lack of concrete models, it is not yet possible to determine the corresponding contribution to the cosmological constant. In the causal set example, something similar to \eqref{EnergyCreationCS} is likely to hold also before decoupling, and thus may largely enhance the corresponding contribution to the effective cosmological constant. In addition, diffusion of non-relativistic particles \cite{Dowker:2003hb} leads to creation of energy of the same form as \eqref{EnergyIncreaseCSL}. For that reason, we did not include a detailed analysis here.
	
	Finally, a very important feature of energy non-conservation in the context of unimodular gravity is that an effective cosmological constant accessible to observations like \eqref{DarkEnergyEstimate1} does not require strong modifications of the local physics. To see this let us consider for simplicity a universe made only of baryons (undergoing spontaneous localization) and photons. Let us assume moreover that the kinetic energy \eqref{EnergyIncreaseCSL} created through CSL is mostly transferred to photons (because of equipartition theorem, and the large number of photons). The back-reaction on the stress-energy tensor is given by the modified continuity equation for photons $\dot \rho^\gamma + 4H\rho^\gamma = \xi_{\va {\va \text{CSL}}} \rho^b$. A solution of this equation in the radiation-dominated era can be written explicitly
\be
\rho^{\gamma}(a)=\rho^{\gamma}_\text{h} \frac{a_\text{h}^4}{a^4}\left[1+\frac{1}{2}\frac{\rho^b_\text{h}}{\rho^\gamma_\text{h}} \frac{\xi_{\va {\va \text{CSL}}}}{H_\text{h}} \left(\frac{a^3}{a_\text{h}^3}-1\right)\right]^{\frac{2}{3}},
\ee
where the subscript $\text{h}$ denotes for the cosmological quantities at hadronization time. The departure from the standard equation of state at the end of radiation-dominated era ($z_\text{eq}\sim 3000$) can read from the quantity
\be
\frac{\rho^\gamma_\text{eq} a_\text{eq}^4}{\rho^\gamma_\text{h} a_\text{h}^4 }-1 \approx \frac{\Omega^\text{b}_0}{2(z_\text{eq} \Omega^\text{m}_0)^{3/2}} \frac{\xi_{\va \text{CSL}}}{H_0} < 10^{-17},
\ee
which is found to be completely negligible. Physically, the above result shows that the energy created, or lost, produces effects that pile up in $\Lambda^\text{eff}$, while their  backreaction on ordinary matter decreases together with the expansion of the universe.
\\

	The computations of the contribution to the effective cosmological constant performed for two models (continuous spontaneous localization and causal sets) illustrate how, despite the smallness of the modifications of the dynamics at the local level, the effect on cosmological scale can be of the order of $\Lambda^\text{obs}$. Moreover, it is  worth noting that, although the quantitative estimates in this paper have been obtained for some specific examples, the results  of our analysis are far more  general, and remain valid as long as the violation of energy conservation is of integrable type \eqref{IntegrableType}. This framework could therefore be used to rule out non-standard models that would lead to an effective cosmological constant that varies to much at late time.
	
There is, a priori,  no reason for the energy momentum violations produced by 
the type of mechanisms evoked here (or those from other hypothetical fundamental sources) 
to satisfy the integrability condition in a general situation. In  cases  where that condition is  violated,  a semiclassical account of  
phenomenon in terms of a metric variable theory of gravity would simply not be viable.  However, in the cosmological setting considered  here,  the cosmological principle---homogeneity and isotropy of the universe at large scales---constraints the 
current $J$ to be of the form $J_t(t)dt$ for which \eqref{IntegrableType} is automatically satisfied at the relevant scales,   making  the framework of unimodular gravity useful despite   possible short scale break in the integrability requirement. 
This, together with the fact that deviations of energy momentum conservation are strongly constrained in local experiments, is what gives phenomenological relevance to our analysis that could also be applied to other situations whenever  the integrability condition can be argued to be approximately valid. 
In more general situations, a metric formulation (seen here as en effective description) would be precluded, and a more fundamental description would need to be found.

	
{ It is however  interesting to point out that the breaking of diffeomorphims invariance down to volume-preserving diffeomorphisms (so that \eqref{IntegrableType} is satisfied down to local scale)  is actually generic in the regime of validity of QFT in curved spacetimes. Concretely, the renormalization of the expectation value of the energy-momentum tensor requires the subtraction of ultra-violate divergences which leads to a normal-ordered stress tensor satisfying
\be\label{anomaly}
	\nabla^a\langle T_{ab} \rangle_{\va {\rm NO}}=\nabla_b Q,
\ee 
where $Q$ is a geometric, state-independent, quantity \cite{Wald:1995yp} (for a simple proof of this fact in 2d see \cite{Fabbri:2005mw}). The standard view, motivated by  consistency with semiclassical general relativity, is to enforce energy-momentum conservation through the redefinition $\langle \tilde T_{ab} \rangle \equiv \langle T_{ab} \rangle_{\va\rm  NO}-Qg_{ab}$. In the case of conformally coupled theories, this leads to the famous trace anomaly, interpreted as a breaking of scale invariance by quantum effects. We can instead simply deal with $\langle T_{ab} \rangle_{\va\rm  NO}$ in the context of unimodular gravity, the physical implications will be the same.  Even if the contributions to the cosmological constant in that case would be tiny, it constitutes a clear-cut example where the type of phenomenon considered here stem from standard quantum effects.\\




	To conclude, we have shown that violation of energy-momentum conservation can be reconciled with metric theory of gravity by taking the fundamental theory  of spacetime to  be  unimodular  gravity. This change of paradigm leads to an effective cosmological constant term in Friedmann's equation, that can be seen as a record of the energy-momentum non-conservation during the history of the universe. It decreases or increases in time, whenever energy is created or lost, yet it becomes quickly a constant (at least in the models described here) as regular matter density dilutes with the expansion.

\section*{Acknowledgments}
	 We  acknowledge    useful  discussions  with    Y. Bonder, L. Diosi,   G. Ellis, M. Knecht,  S. Lazzarini, J. Navarro-Salas, P. Pearle,  M. Reisenberger,  S. Saunders,  and A. Tilloy. We thank Seth Major for pointing out the relevance of the causal set diffusion models.  We are also grateful to an anonymous referee who contributed to improving the presentation of our work.  DS  acknowledges partial financial support from DGAPA-UNAM project IG100316 and  by CONACyT project 101712.  AP acknowledges the OCEVU Labex (ANR-11-LABX-0060) and the A*MIDEX project (ANR-11-IDEX-0001-02) funded by the``Investissements d'Avenir" French government program managed by the ANR.


\end{document}